\documentclass[10pt, conference]{IEEEtran}
\usepackage{enumerate}
\usepackage{booktabs}
\usepackage{multirow}
\usepackage{algorithm}
\usepackage{algpseudocode}
\usepackage{comment}
\usepackage{amsmath,amssymb}
\usepackage{latexsym}
\usepackage{graphicx}
\usepackage{subfigure}
\usepackage{epstopdf}
\usepackage{epsfig}
\usepackage{xspace} 
\usepackage[bottom]{footmisc}
\usepackage[11pt]{moresize}
\usepackage{url}
\newcommand{\mypara}[1]{
	\vspace*{0.01cm}
	\noindent\textbf{\textit{#1}}}

\begin{document}
\title{PVF: Understanding AI Vulnerability Against SDCs}
\author{Xun Jiao, Fred Lin, Harish Dixit, Joel Coburn, Sajin Nair, Abhinav Pandey \\ Han Wang, Venkat Ramesh, Jianyu Huang, Daniel Moore, Sriram Sankar \\ Meta Platforms, Inc}
\maketitle

\begin{abstract}
Reliability of AI systems is a fundamental concern for the successful deployment and widespread adoption of AI technologies. Unfortunately, the escalating complexity and heterogeneity of AI hardware systems make them increasingly susceptible to hardware faults, e.g., silent data corruptions (SDC), that can potentially corrupt model parameters. When this occurs during AI inference/servicing, it can potentially lead to incorrect or degraded model output for users, ultimately affecting the quality and reliability of AI services. In light of the escalating threat, it is crucial to address key questions: How vulnerable are AI models to SDCs, and how do different parts (such as modules and layers) of the models exhibit varying vulnerability levels to SDCs? To answer this question, we propose a novel quantitative metric, Parameter Vulnerability Factor (PVF), inspired by architectural vulnerability factor (AVF) in computer architecture community, aiming to standardize the quantification of AI model vulnerability against SDCs. We define a model parameter's PVF as the probability that a corruption in that particular model parameter will result in an incorrect output. Similar to AVF, this statistical concept can be derived from statistically extensive and meaningful fault injection (FI) experiments. 
In this paper, we present several use cases on applying PVF to three types of tasks/models during inference -- recommendation (DLRM), vision classification (CNN), and text classification (BERT), while presenting an in-depth vulnerability analysis on DLRM. In DLRM, our FI results show that different parts of DLRM present different vulnerability levels: top-MLP layers are the most vulnerable parameter component, while embedding tables exhibit comparatively lower vulnerability level. PVF has been a critical metric used for making key error management design decisions in productionizing Meta's in-house AI chip - MTIA~\cite{coburn2025meta}.  

\end{abstract}

\section{Introduction}
The reliability of AI systems directly translates to the dependability, safety, and functionality of services running on top of them. For instance, in recommendation model inference, a reliable model is essential for accurate personalized recommendations, crucial for achieving positive business outcomes.  
Unfortunately, as AI hardware systems become more complex and heterogeneous, and as transistor technology plunges into the deep-nanometer regime, the reliability of AI hardware systems faces a mounting challenge and a rising susceptibility to faults that can potentially corrupt model parameters. One pronounced threat that has been gaining increasing attention recently in hardware systems is data corruption, referring to errors or alterations in data that may occur during storage, transmission, or processing, leading to unintended changes in information. This can happen due to manufacturing defects, aging components, or environmental factors~\cite{hazucha2003neutron, sangchoolie2017one, agarwal2023resilience, chang2018evaluating}. For AI systems, these faults can impact accuracy, integrity, and reliability of high-level AI application quality. 

In particular, hardware faults that are not reported by standard fault reporting mechanisms but leading to erroneous application behavior have become increasingly prominent and harder to detect. We refer to these as \textbf{silent data corruption (SDC)}, and has been reported by Meta~\cite{dixit2022detecting}, and confirmed by Google and Alibaba~\cite{hochschild2021cores, wang2023understanding}. 
In AI systems, Nvidia reported that ``Hopper architecture GPUs may intermittently experience SDC resulting in incorrect results''~\cite{nvidia-sdc}, and Google reported hard to debug SDCs in their Tensor Processing Unit (TPU) systems~\cite{he2023understanding}.

\begin{figure*}[htbp!]
    \centering
    \includegraphics[width=1.5\columnwidth]{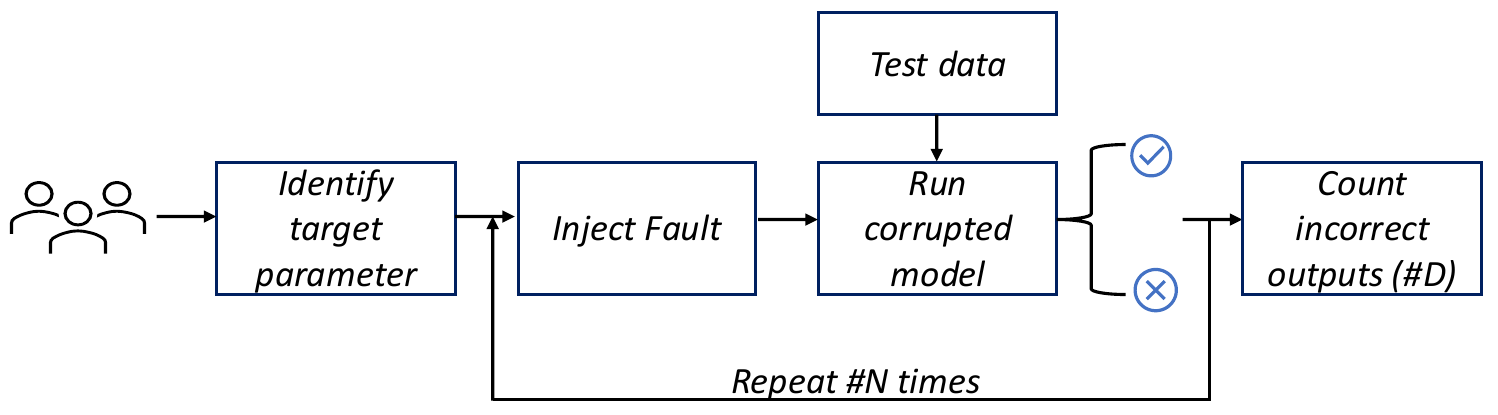}
    \caption{Fault injection experiments flow}
    \label{fig:fi}
\end{figure*}

Existing work on evaluating AI vulnerability against hardware faults used metrics such as accuracy drop~\cite{reagen2018ares, mahmoud2020pytorchfi} or SDC rate~\cite{li2017understanding, agarwal2023resilience}, focusing on model-level vulnerability, but there is a lack of an unified parameter-level vulnerability metric that can answer this question: \textit{How likely is a parameter corruption to result in an incorrect model output?} The answer to this question is critical in AI hardware design, especially when mapping AI model parameters or software variables to hardware blocks which may have varying fault protection capabilities. 
Thus, we propose the PVF metric, with the following features: 
\begin{itemize}
    \item Parameter-level Quantitative Assessment: As a quantitative metric, PVF concentrates on parameter-level vulnerability, calculating the likelihood that a corruption in a specific model parameter will lead to an incorrect model output. This ``parameter'' can be defined at different scales and granularities, such as an individual parameter or a group of parameters. By applying PVF to several example models, we quantify the varying vulnerability levels of different parts of a given model.
    \item Scalability Across AI Models/Tasks: PVF can be scalable and applicable across a wide range of AI models, tasks, and hardware fault models. While in this paper, we present case studies on three types of AI classification tasks/models (recommendation, vision, and text), PVF can be applied to other AI models with model-specific attributes such as the definition of ``incorrectness''. We further discuss potential extension of PVF to training.
    \item Guiding AI System Design: PVF can provide pivotal insights for AI system designers, guiding them in making informed decisions about balancing fault protection with performance and efficiency, e.g., map higher vulnerable parameters to better-protected hardware blocks and explore tradeoffs on latency, power and reliability by enabling a surgical approach to fault tolerance at selective locations instead of a ``catch-all/none'' approach. 
    \item Standardization of AI Vulnerability/Resilience Evaluation: We propose to use PVF as a standard metric for AI vulnerability/resilience evaluation. It has the potential to unify and standardize such practices, making it easier to compare the reliability of different AI systems/parameters and fostering open collaboration and progress in the industry/research community.
\end{itemize}
PVF has been a critical metric used for making key error management design decisions in productionizing Meta's in-house AI chip - MTIA~\cite{coburn2025meta}. Specifically, we used PVF to quantify and identify which parts of a model are most sensitive to errors and how to mitigate them. We found that bit flips in Table Batched Embedding (TBE) indices, TBE table rows, or specific bits in floating-point representations of dense weights can cause NaNs or output corruptions, with some failures occurring with high probability.

\section{Related Work}
Evaluating AI vulnerability against hardware faults has been a topic of increasing focus and priority with recent SDC findings from hyperscalars such as Meta~\cite{dixit2022detecting}~\cite{reagen2018ares, mahmoud2020pytorchfi, li2017understanding}), not only because of increasing susceptibility of AI systems to hardware faults in general, but also the increasing need to improve AI hardware acceleration by balancing the tradeoff between latency and fault protection. For example, Ares~\cite{reagen2018ares} and PyTorchFI~\cite{mahmoud2020pytorchfi} are two popular tools on top of PyTorch that can inject faults to AI models and evaluate the resulting accuracy drop. In addition, another widely-used evaluation metric is SDC rate which compute the probability of output corruption by performing FI experiments. For instance, Li et al. compute the SDC rates of CNNs~\cite{li2017understanding} and Agarwal et al. compute the SDC rates for large language models~\cite{agarwal2023resilience}. 

However, these evaluation metrics are focused on model-level vulnerability; there is a lack of an unified metric to quantify parameter-level vulnerability. Quantification of model parameter vulnerability has much implication in AI system software-hardware codesign, especially when mapping software parameters to hardware blocks with different fault protection capabilities. For example, given the different fault management capabilities of SRAMs, DRAMs and HBMs, there are varying degrees of vulnerabilities under consideration while mapping and allocating different model parameters. 
To facilitate scenarios like this and unify evaluation across AI models and fault scenarios, we propose PVF. 

Another set of relevant work is in the field of adversarial attack~\cite{sun2021exploring, he2019parametric, rakin2019bit}; these works use gradient-based methods looking for most vulnerable parameter. Our work is fundamentally different from these works because our target is different, and fault model is different (these works often assume statistical corruption such as Gaussian noise which cannot handle faults such as NaN due to bit flips).

\section{Computing PVF}
\subsection{PVF Definition}
PVF is inspired by architectural vulnerability factor (AVF) in computer architecture community. AVF quantifies the vulnerability of a processor's microarchitecture to soft errors~\cite{mukherjee2003systematic}. An architectural structure's AVF is the probability that a fault in that particular structure will result in a program output error. Similarly, we define a model parameter’s PVF as the probability that a corruption in that particular model parameter will result in an incorrect model output, e.g., a wrong click prediction or a wrong image classification. Note that PVF can scale to different fault models and parameter granularities, which is shown later. 
As a statistical concept, PVF needs to be derived through a large number of fault injection (FI) experiments that are statistically meaningful.

\subsection{Fault Model}
\label{sec:fm}
In this paper, we consider parameter corruptions using three hardware fault models. These simulate hardware faults such as memory bit flips due to soft errors~\cite{hazucha2003neutron}. These fault models are used in previous studies~\cite{reagen2018ares, mahmoud2020pytorchfi, kim2018matic}. We recognize that different hardware platforms exhibit different error patterns, and the fault model used here may not fully represent realistic hardware error patterns. However, PVF is adaptable to any user-specified fault model based on users' specific scenarios.  

\mypara{Single-Bit Flip (SBF):} Under SBF model, we inject a random single bit flip to the target parameter component (e.g., embedding table) during each FI experiment. This is the most widely used fault model across different resilience studies~\cite{sangchoolie2017one, agarwal2023resilience, chang2018evaluating}. 

\mypara{Multiple Bit Flip (MBF)/Bit Error Rate (BER):} Under MBF model, we inject multiple bit flips to the target parameter component during each FI experiment. For example, instead of injecting 1 bit flip, we can inject 32 bit flips to the embedding table. Multiple bit flips can be injected either using a fixed number of multiple bits, or using a bit error rate (BER), which is the portion of the bits getting flipped based on the target parameter count. These two metrics are technically interchangeable and one can be derived using the other. 
In our FI experiment, we decide to use the fixed-number based multiple bit flip model. This is to ensure that target parameter with small parameter count (e.g., 48 weights in the $8^{th}$ embedding table of the experimented DLRM) under application of a realistic BER (e.g., $10^{-10}$) does not yield a practically unusable number for the FI experiment.

\mypara{Multiple Burst Bit Flip (MBBF): } Under MBBF model, we assume a burst error model, where for each FI experiment, two consecutive bits will get flipped.

\subsection{Fault Injection (FI) Approach}
We use the following steps to perform FI experiments for any given AI model during inference (Fig.~\ref{fig:fi}):
\begin{itemize}
    \item Step 1 Load the (pre-trained) AI Model: Since we focus on inference for now, we assume there is a pre-trained AI model already. We load the pre-trained AI model (weights) that we want to inject faults. 
    \item Step 2 Identify Target Parameters: We identify the specific parameters in the AI model that we want to compute PVF; examples include a specific embedding table, a specific convolutional kernels, or a specific fully-connected layer, or other relevant parts of the model.
    \item Step 3 Inject Faults: For the target parameter component, we inject faults based on the given fault model. For example, if using SBF for an embedding table with 100 embedding weights where each weight has 32 bits, we flip 1 bit randomly sampled from the 3200 bits. 
    \item Step 4 Evaluate the Corrupted Model: Run the AI model inference with the injected faults on the test data, and compare the model's output with groundtruth output. Step 3-4 is considered as one FI experiment. 
    \item Step 5 Repeat for $N$ times: Repeat Step 3-4 for $N$ times (e.g., 1 million) wherein each FI we use different random faults and inputs, and record the number of incorrect output ($D$). 
    \item Step 6 Calculate PVF: Compute PVF as the $D/N$. Note, because the model can have wrong predictions without any hardware errors, we only focus on cases where correct predictions become incorrect. 
\end{itemize}

\section{Case Study on DLRM}
\subsection{DLRM Architecture}
\begin{figure}[htbp!]
    \centering
    \includegraphics[width=0.99\columnwidth]{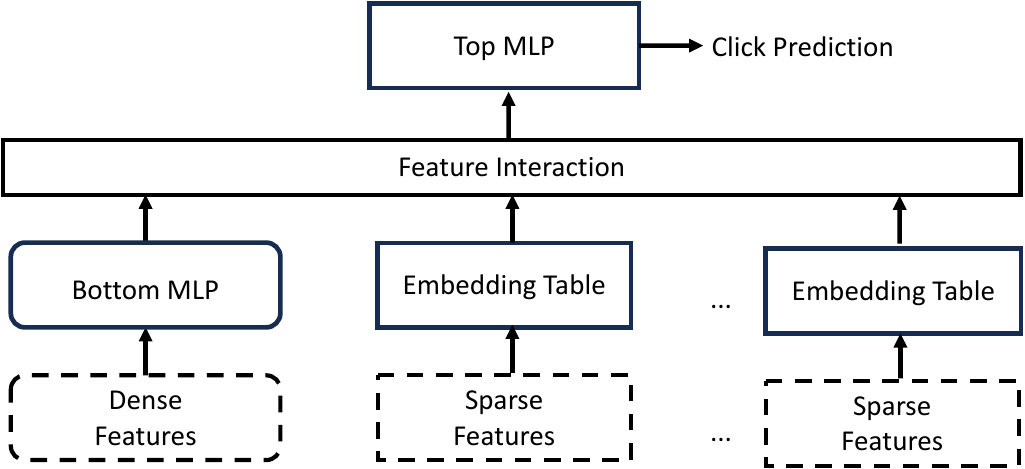}
    \caption{DLRM architecture overview}
    \label{fig:dlrm_overview}
\end{figure}
We present the first case study on DLRM inference. DLRM was developed by Meta for personalized content recommendation, and has constituted 79\% of the overall AI inference cycles at the Meta data center~\cite{hsia2023mp}. The fundamental architecture of a DLRM is shown in Fig.~\ref{fig:dlrm_overview}, composed of three key parameter components: embedding tables, bottom MLP (bot-MLP), and top MLP (top-MLP). Embedding tables transform sparse categorical features into dense, continuous representations. These embeddings capture latent features and relationships, allowing the model to discern intricate patterns and correlations in the data. 
The MLP layers in DLRM contains a bot-MLP and a top-MLP. Firstly, the dense features undergo transformation through the bot-MLP. This segment comprises a sequence of linear layers accompanied by Rectified Linear Unit (ReLU) activations. Subsequently, the outcome from the bot-MLP and the embedding vectors are combined in a feature interaction such as feature concatenation. The result of this interaction is then fed into the top-MLP. The top-MLP generates the final model output, e.g., click prediction. 

We focus on fault injection to model parameters such as MLP layers and embedding tables because they make most of the total storage needed for a model. Further, they are essentially read-only and static so bit flips are likely to have persisting impact, whereas for intermediate model data such as activations, they have short lifetime before they are updated again. Other data structures critical to model execution will typically be protected, similar to protecting stack/heap, etc., so we do not consider these.

\subsection{Experimental Setup}
We train the \textit{DLRM} using Criteo DAC dataset~\cite{criteo-dac}, as used by the original DLRM paper, and achieve the baseline accuracy (78.83\%), aligned with the original DLRM paper~\cite{naumov2019deep}. DLRM uses 32-bit floating point number (FP32) as the default data type for its parameters. 
The Criteo-DAC dataset contains the click records of 45 million users over a 7-day span with 13 dense features and 26 categorical features. 
The number of parameters of embedding tables ranges between 48 to 162 million, bot-MLP and top-MLP components have 155984 and 320001 parameters, respectively. 
To be statistically meaningful, each PVF is derived by performing 1 million independent FI experiments. 

\subsection{PVF under MBF}
\begin{figure}[htbp!]
    \centering
    \includegraphics[width=0.9\columnwidth]{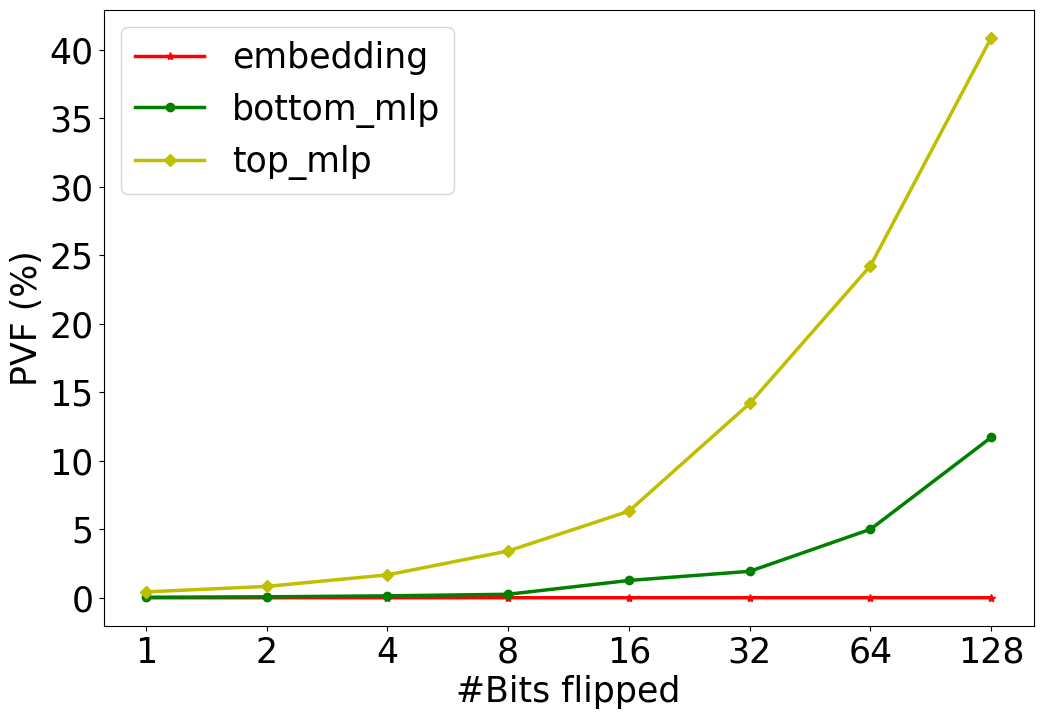}
    \caption{PVF of DLRM Parameters under MBF}
    \label{fig:pvf_multi_bit}
\end{figure}
Fig.~\ref{fig:pvf_multi_bit} illustrates the PVF of three DLRM parameter components, embedding table, bot-MLP, and top-MLP, under MBF fault models with 1, 2, 4, 8, 16, 32, 64, and 128 bit flips during each inference. We can observe several important facts. 

\mypara{Observation 1: Different parts of DLRM present different vulnerability levels.}
We observe that different parts of DLRM present different vulnerability level, e.g., under a single bit flip, most embedding tables have low PVF, e.g., less than 0.0001\%; however, top-MLP can have 0.4\% under even a single bit flip. This is significant -- for every 1000 inferences, 4 inferences will be incorrect. This highlights the importance of protecting specific vulnerable parameters for a given model based on the PVF measurement. 


We observe that even with 128 bit flips during each inference, the PVF of embedding table is still low ($<0.0001\%$), exhibiting notable resilience against corruptions; for MLP components, PVF has increased to 40\% or 10\% for top-MLP and bot-MLP components respectively, while observing multiple NaN values.  
This is attributed to embedding tables being highly sparse, and parameter corruptions are only activated when the particular corrupted parameter is ``hit'' by the corresponding sparse feature. 
Even if activated, it may get masked by the subsequent neural processing. Consequently, a bit flip in the embedding table is likely to not get exposed or reflected in the DLRM output in most cases. This further lowers the probability of error exposure for the architecture under study.
Note that, because different parameter components has different parameter count, the same bit flip count is essentially equivalent to different BERs for these parameter components. For example, 1000 bit flips for embedding table is 0.000185\%, and for top-MLP and bot-MLP are 0.641\% and 0.3124\%. Thus, when we compare the PVF across different parameter components, we need to explicitly state the underlying fault model (which can be specified by users based on the assumption regarding hardware faults) to make a fair and clear comparison. 

\begin{figure}[htbp!]
    \centering
    \includegraphics[width=0.9\columnwidth]{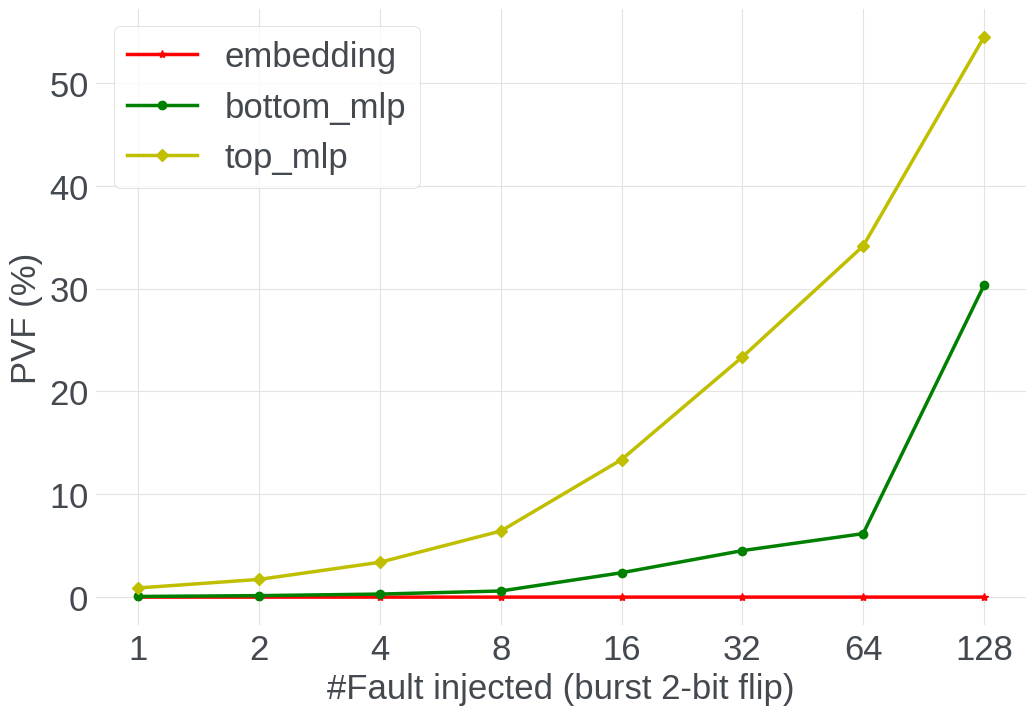}
    \caption{PVF of DLRM Parameters under MBBF}
    \label{fig:pvf_multi_burst_bit}
\end{figure}

Fig.~\ref{fig:pvf_multi_burst_bit} presents the PVF under burst errors, where we sweep the burst error counts from 1 to 128. In this case, each burst error will flip two consecutive bits. Results show that, even at 128 burst errors, the PVF of embedding table is still low ($<0.0001\%$), while for top mlp and bottom mlp, the PVF increased to 50\% and 30\% respectively under 128 burst errors, higher than the PVF under MBF which is intuitively reasonable.

\subsection{PVF under SBF}
\begin{figure}[htbp!]
    \centering
    \includegraphics[width=0.9\columnwidth]{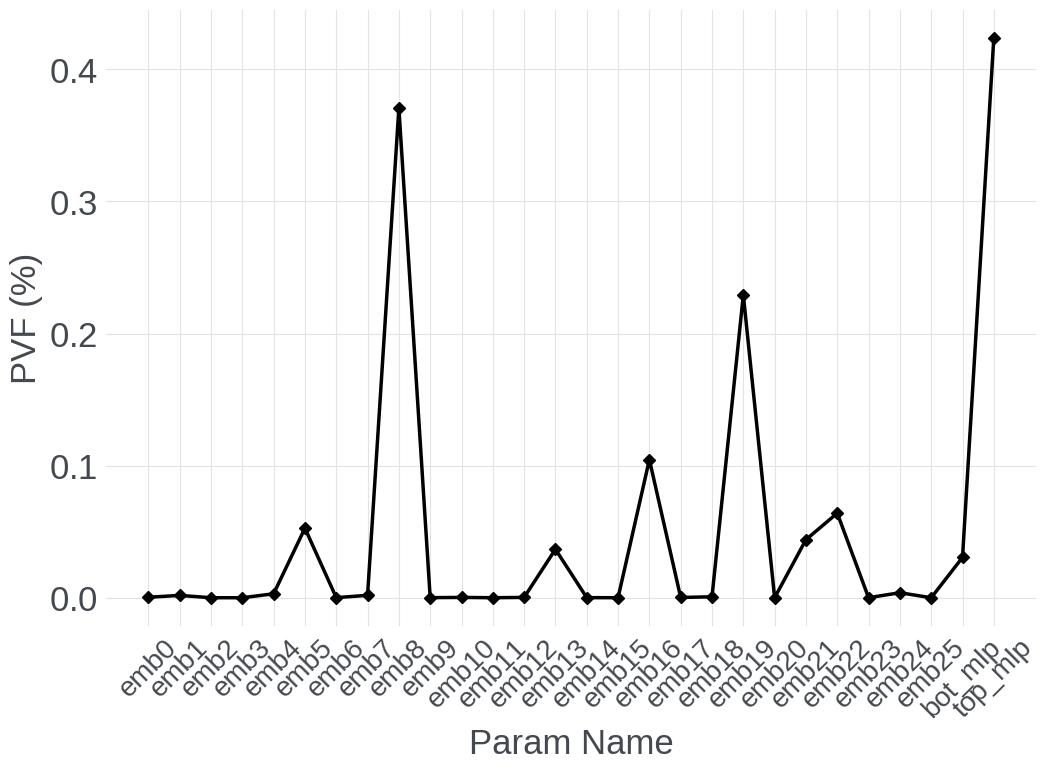}
    \caption{PVF of DLRM Parameters under SBF}
    \label{fig:pvf}
\end{figure}

To further break down the granularity of target parameters, we measure the PVF of individual embedding tables \textit{under the SBF fault model}. The results are shown in Fig.~\ref{fig:pvf}. 

\begin{figure}[htbp!]
    \centering
    \includegraphics[width=0.9\columnwidth]{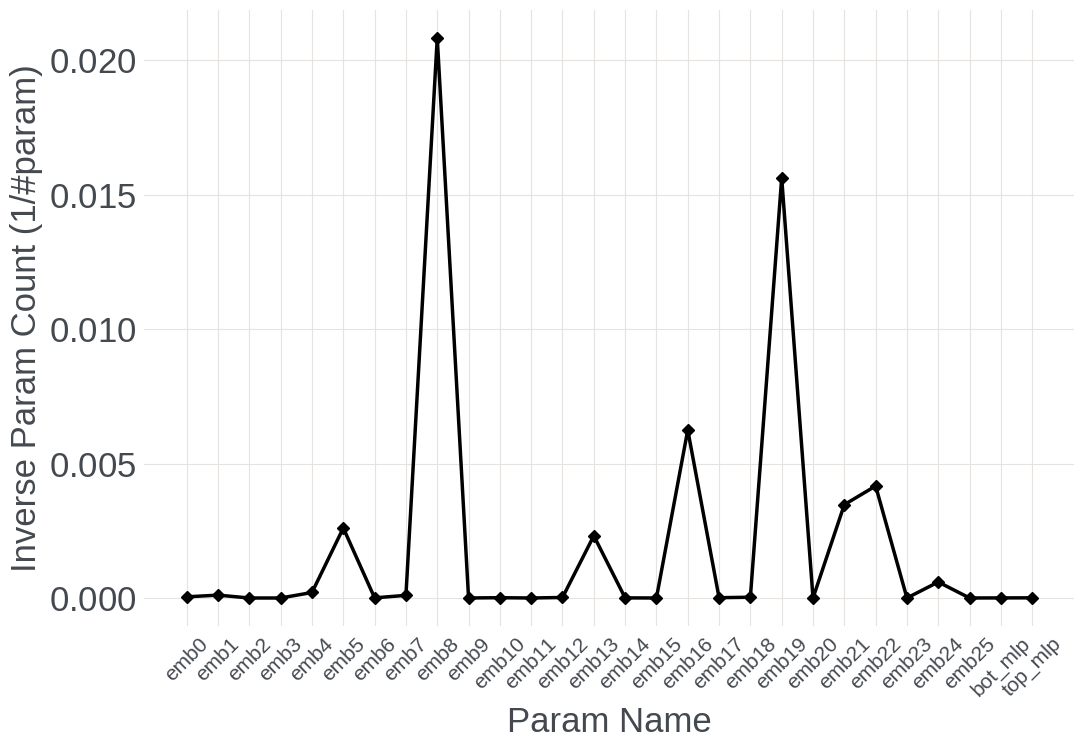}
    \caption{Inverse of Number of DLRM Parameters (1/count)}
    \label{fig:count}
\end{figure}

\mypara{Observation 2: Top-MLP has higher vulnerability than bot-MLP and embedding tables, while different individual embedding tables show different vulnerabilities. }
Based on Fig.~\ref{fig:pvf}, we observe that MLP components have higher PVF than embedding tables. The top-MLP component has highest PVF among all target parameter components, while bot-MLP component has higher PVF than most embedding tables. The reason for this is that embedding tables are highly sparse (stated in Observation 2 and 3 earlier).  
Consequently, a bit flip in the embedding table is likely not to get exposed or reflected in the DLRM processing in most cases. This further lowers the probability of error reflection.

We also observe that top MLP component has higher PVF than bottom MLP. This is attributed to the top-MLP being closer to the final model, and hence has less of a chance to be mitigated by inherent error masking probability of DLRM models. Note that in this case, top-MLP has more parameters than bot-MLP -- meaning that same bit flip count suggests smaller BER for top-MLP. 

Under SBF fault model, we observe that the PVF of embedding tables varies by orders of magnitude, and is highly (inversely) correlated with the parameter count (as seen in Fig.~\ref{fig:count}), which shows the inverse of the parameter count. We draw a correlation that, the smaller an embedding table is, the higher its PVF (under SBF model). This is intuitively reasonable because under SBF fault model, a single bit flip will always occur in the target parameter regardless of the size of the target parameter. Thus, for an embedding table with smaller parameter count, it is more likely that the bit flip will be activated by the sparse feature.  
However, it is worth to note that, for a smaller embedding table, the probability of a bit flip occurrence is also smaller than a larger embedding table.

\begin{figure}[htbp!]
    \centering
    \includegraphics[width=0.9\columnwidth]{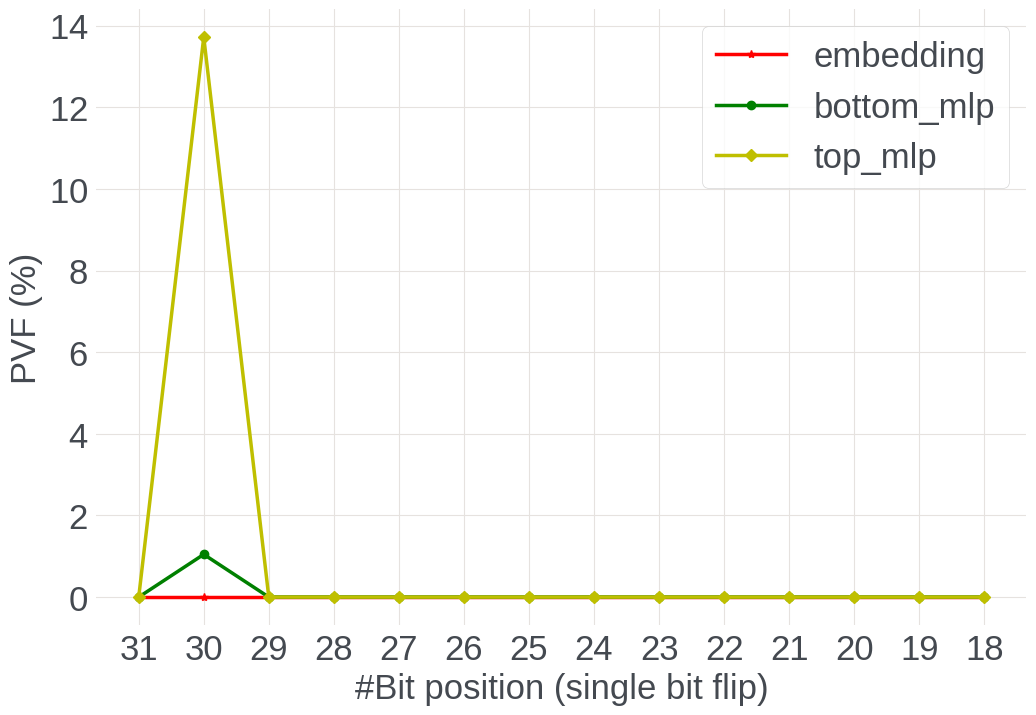}
    \caption{PVF of different bit positions}
    \label{fig:bit_pos}
\end{figure}

\mypara{Observation 3: Different bits have different vulnerability levels.} 
Fig.~\ref{fig:bit_pos} presents PVF of different bit positions (we omit some least significant bits because they consistently have 0 PVF) for embedding, top-MLP, and bot-MLP. In this scenario, to obtain the PVF of a given bit position, we only inject bit flips to that specific bit position. A key observation is that the sign bit (bit 31) is not the most vulnerable bit. Bit 30 is the most vulnerable bit because compared to other bits, it is more likely to result in large values as well as abnormal data such as NaNs; in FP32 data, NaN could be caused by a ``0'' to ``1'' flip at bit position 30.

\section{Case Study on CNN Models}
We present the second case study on convolutional neural networks (CNNs) for vision classification tasks. For the sake of simplicity, we pick a small CNN, LeNet~\cite{lecun1998gradient}, for MNIST dataset~\cite{lecun1989handwritten}, under the SBF model. We show the CNN layers and the corresponding weights count in Table~\ref{tab:feature}.

\begin{table}[htbp!]
\caption{CNN layers w/ weight counts}
\centering
    \begin{tabular}{ccc}
    \toprule
    Layer Name       & Weight Size & Weight Count  \\
    \midrule
    conv1       &  [6, 1, 5, 5]    & 150   \\
    conv2       &  [16, 6, 5, 5]   & 2400       \\
    fc1         &  [120, 256]      & 30720 \\
    fc2         &  [84, 120]       & 10080 \\
    fc3         &  [10, 84]        & 840 \\
    \bottomrule  
    \end{tabular}
  \label{tab:feature}
\end{table}

\begin{figure}[htbp!]
    \centering
    \includegraphics[width=0.9\columnwidth]{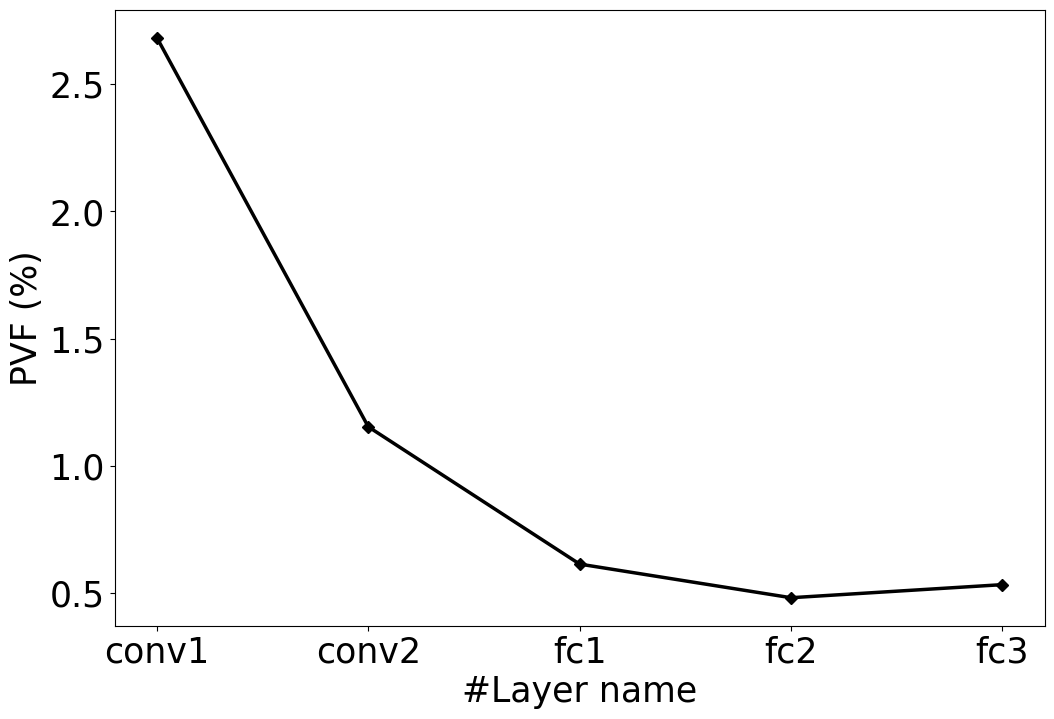}
    \caption{PVF of LeNet under SBF}
    \label{fig:lenet}
\end{figure}

We perform 100K FI experiments for each target parameter, and present the obtained PVF for each layer of LeNet in Fig~\ref{fig:lenet}, where we can see that, conv1 has the highest PVF among all the conv (convolutional) and fc (fully-connected) layers. That is, a single bit flip in conv1 is more likely to result in an incorrect output. However, it is worth to note that, similar to previous discussion, the weight count of conv1 is only 0.06\% of conv2 and 0.488\% of fc1, thus the probability of having a bit flip in conv1 is less than conv2 and fc1 under same hardware condition. A fairer comparison would be using BER for all parameters; however, for a small BER such as $10^{-7}$, conv1 layer will not even produce a single bit flip for most FI experiments.

\section{Case Study on NLP Models}
We also present a case study on applying PVF to NLP models. For the sake of simplicity, we fine-tuned a smaller BERT model~\cite{devlin2018bert} with four encoder layers, for ag-news dataset~\cite{zhang2015character}, under the SBF model. For the experimental setup, we directly use the torchtext.datasets() utility function to load the ag-news dataset. We target 12 BERT parameter components: for each encoder layer from layer 0 to layer 3, we target query.weight, key.weight, and value.weight, which are the weights that perform linear transformations to the original input token. Each weight component has same shape ([256, 256]) and 65536 parameter counts.

\begin{figure}[htbp!]
    \centering
    \includegraphics[width=0.9\columnwidth]{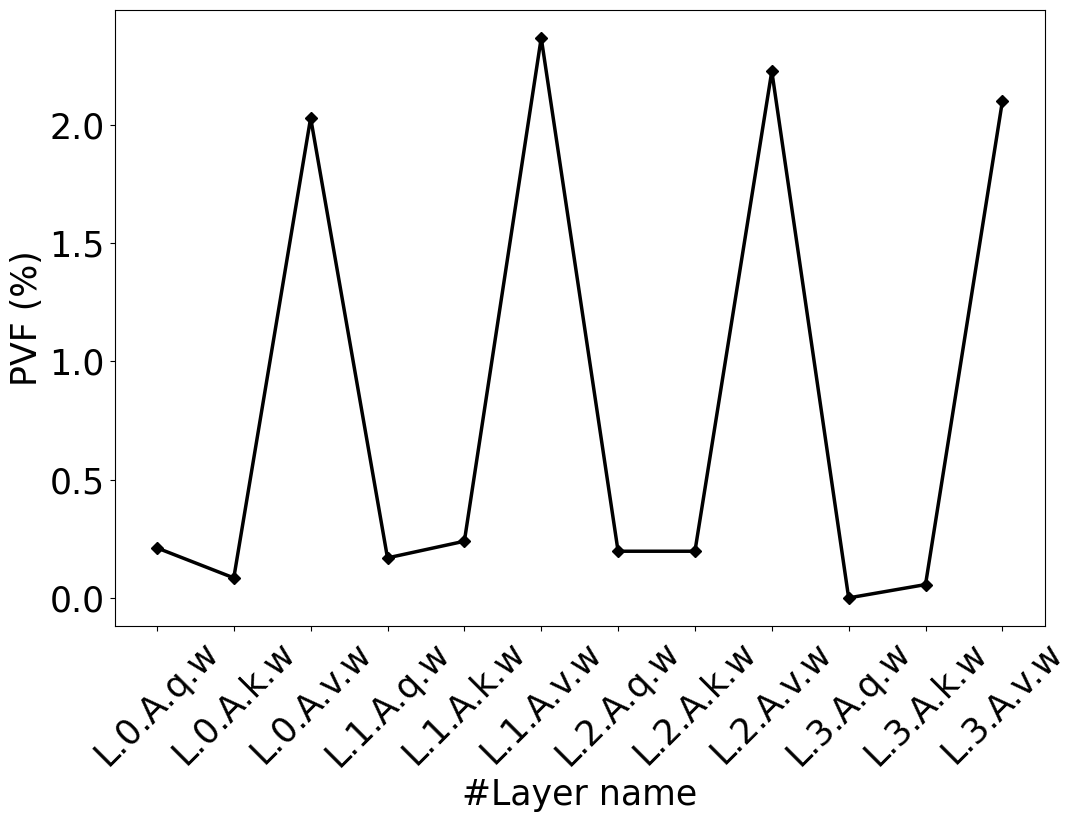}
    \caption{PVF of Tiny BERT under SBF. (L.0.A.q.w means layer.0.attention.query.weight)}
    \label{fig:bert}
\end{figure}

We perform 100K FI experiments for each target parameter, and present the obtained PVF for each target component in Fig~\ref{fig:bert}, where we can see that, value.weight parameter components always have the highest PVF among all the components, regardless of its specific layer. This is intuitively reasonable because in the original self-attention equation, both $Q$ (query) and $K$ (key) would go through more transformations such as downscaling (divide by $\sqrt{d_k}$) and softmax which may mitigate certain bit flips, while $V$ almost directly determines the self-attention output. 

\begin{equation}
Attention(Q, K, V) = \text{softmax}\frac{Q \cdot K^T}{\sqrt{d_k}} \cdot V
\end{equation}

\section{Discussion}
PVF is a versatile metric that can be tailored to various AI models/tasks. The definition of an ``incorrect output'' will vary based on the model/task and can be adapted to suit user requirements. For instance, in a large language model (LLM) used for question-answering tasks, an incorrect output would be an incorrect answer. For a GenAI model/task, users might have a specific metric to define ``incorrectness'' for the generated content. We anticipate that the introduction of PVF will stimulate diverse use cases in both research and production settings. 

PVF is also adaptable to various hardware fault models. While our study presents PVF under three fault models, we acknowledge that different hardware platforms under varying physical conditions may exhibit different fault models, such as bit flips, statistical variations, etc. The principle of PVF remains applicable, and the method to calculate PVF remains consistent. The only modification required is the manner in which the fault is injected, based on the assumed fault models. 

Furthermore, PVF can be extended to the training phase to evaluate the effects of parameter corruptions on the model's convergence capability. During training, the model's parameters are iteratively updated to minimize a loss function. A corruption in a parameter could potentially disrupt this learning process, preventing the model from converging to an optimal solution. By applying the PVF concept during training, we could quantify the probability that a corruption in each parameter would result in such a convergence failure. A key distinction between training and inference is that during training, the parameters are dynamically updated, making the PVF potentially a function of time as well. We reserve this aspect for future exploration and investigation.

\section{Conclusion}
In this research, we propose a novel quantitative metric, the Parameter Vulnerability Factor (PVF), designed to quantify the vulnerability of AI models to parameter corruptions. Through fault injection, PVF can be calculated for any target parameter component of a given AI model. In this paper, we present three case studies where we apply PVF to recommendation (DLRM), vision classification (CNN), and text classification (BERT). The PVF measurements allow us to examine the vulnerability levels of different parameter components within a specific model. The introduction of the PVF metric provides crucial insights for AI hardware designers, assisting them in balancing fault protection with performance and efficiency. For instance, it can guide the assignment of vulnerable AI parameter components to highly protected hardware modules. The PVF metric is versatile and can be applied to any AI model. It has the potential to serve as a unifying and standardizing tool for evaluating AI vulnerability and resilience.
\bibliographystyle{plain}
\bibliography{main}

@inproceedings{coburn2025meta,
  title={Meta's Second Generation AI Chip: Model-Chip Co-Design and Productionization Experiences},
  author={Coburn, Joel and Tang, Chunqiang and Asal, Sameer Abu and Agrawal, Neeraj and Chinta, Raviteja and Dixit, Harish and Dodds, Brian and Dwarakapuram, Saritha and Firoozshahian, Amin and Gao, Cao and others},
  booktitle={Proceedings of the 52nd Annual International Symposium on Computer Architecture},
  pages={1689--1702},
  year={2025}
}

@article{zhang2015character,
  title={Character-level convolutional networks for text classification},
  author={Zhang, Xiang and Zhao, Junbo and LeCun, Yann},
  journal={Advances in neural information processing systems},
  volume={28},
  year={2015}
}

@article{lecun1989handwritten,
  title={Handwritten digit recognition with a back-propagation network},
  author={LeCun, Yann and Boser, Bernhard and Denker, John and Henderson, Donnie and Howard, Richard and Hubbard, Wayne and Jackel, Lawrence},
  journal={Advances in neural information processing systems},
  volume={2},
  year={1989}
}

@article{lecun1998gradient,
  title={Gradient-based learning applied to document recognition},
  author={LeCun, Yann and Bottou, L{\'e}on and Bengio, Yoshua and Haffner, Patrick},
  journal={Proceedings of the IEEE},
  volume={86},
  number={11},
  pages={2278--2324},
  year={1998},
  publisher={Ieee}
}

@article{dixit2022detecting,
  title={Detecting silent data corruptions in the wild},
  author={Dixit, Harish Dattatraya and others},
  journal={arXiv preprint arXiv:2203.08989},
  year={2022}
}

@inproceedings{hazucha2003neutron,
  title={Neutron soft error rate measurements in a 90-nm CMOS process and scaling trends in SRAM from 0.25-/spl mu/m to 90-nm generation},
  author={Hazucha, Peter and others},
  booktitle={IEDM},
  year={2003}
}

@inproceedings{reagen2018ares,
  title={Ares: A framework for quantifying the resilience of deep neural networks},
  author={Reagen, Brandon and others},
  booktitle={DAC},
  year={2018}
}

@inproceedings{li2017understanding,
  title={Understanding error propagation in deep learning neural network (DNN) accelerators and applications},
  author={Li, Guanpeng and others},
  booktitle={SC},
  year={2017}
}

@inproceedings{chang2018evaluating,
  title={Evaluating and accelerating high-fidelity error injection for hpc},
  author={Chang, Chun-Kai and Lym, Sangkug and Kelly, Nicholas and Sullivan, Michael B and Erez, Mattan},
  booktitle={SC18: International Conference for High Performance Computing, Networking, Storage and Analysis},
  pages={577--589},
  year={2018},
  organization={IEEE}
}

@inproceedings{agarwal2023resilience,
  title={Resilience Assessment of Large Language Models under Transient Hardware Faults},
  author={Agarwal, Udit Kumar and Chan, Abraham and Pattabiraman, Karthik},
  booktitle={2023 IEEE 34th International Symposium on Software Reliability Engineering (ISSRE)},
  pages={659--670},
  year={2023},
  organization={IEEE}
}

@inproceedings{mukherjee2003systematic,
  title={A systematic methodology to compute the architectural vulnerability factors for a high-performance microprocessor},
  author={Mukherjee, Shubhendu S and Weaver, Christopher and Emer, Joel and Reinhardt, Steven K and Austin, Todd},
  booktitle={Proceedings. 36th Annual IEEE/ACM International Symposium on Microarchitecture, 2003. MICRO-36.},
  pages={29--40},
  year={2003},
  organization={IEEE}
}

@inproceedings{sangchoolie2017one,
  title={One bit is (not) enough: An empirical study of the impact of single and multiple bit-flip errors},
  author={Sangchoolie, Behrooz and Pattabiraman, Karthik and Karlsson, Johan},
  booktitle={2017 47th annual IEEE/IFIP international conference on dependable systems and networks (DSN)},
  pages={97--108},
  year={2017},
  organization={IEEE}
}

@inproceedings{mahmoud2020pytorchfi,
  title={Pytorchfi: A runtime perturbation tool for dnns},
  author={Mahmoud, Abdulrahman and others},
  booktitle={DSN-W},
  year={2020}
}

@inproceedings{kim2018matic,
  title={MATIC: Learning around errors for efficient low-voltage neural network accelerators},
  author={Kim, Sung and others},
  booktitle={DATE},
  year={2018}
}

@inproceedings{he2023understanding,
  title={Understanding and Mitigating Hardware Failures in Deep Learning Training Systems},
  author={He, Yi and Hutton, Mike and Chan, Steven and De Gruijl, Robert and Govindaraju, Rama and Patil, Nishant and Li, Yanjing},
  booktitle={Proceedings of the 50th Annual International Symposium on Computer Architecture},
  pages={1--16},
  year={2023}
}

@inproceedings{hsia2023mp,
  title={MP-Rec: Hardware-Software Co-design to Enable Multi-path Recommendation},
  author={Hsia, Samuel and others},
  booktitle={ASPLOS},
  year={2023}
}

@inproceedings{wang2023understanding,
  title={Understanding Silent Data Corruptions in a Large Production CPU Population},
  author={Wang, Shaobu and Zhang, Guangyan and Wei, Junyu and Wang, Yang and Wu, Jiesheng and Luo, Qingchao},
  booktitle={Proceedings of the 29th Symposium on Operating Systems Principles},
  pages={216--230},
  year={2023}
}

@inproceedings{nvidia-sdc,
  title={tesla-release-notes: https://docs.nvidia.com/datacenter/tesla/tesla-release-notes-535-129-03/index.html},
  year={2023}
}

@inproceedings{sun2021exploring,
  title={Exploring the vulnerability of deep neural networks: A study of parameter corruption},
  author={Sun, Xu and Zhang, Zhiyuan and Ren, Xuancheng and Luo, Ruixuan and Li, Liangyou},
  booktitle={Proceedings of the AAAI Conference on Artificial Intelligence},
  volume={35},
  number={13},
  pages={11648--11656},
  year={2021}
}

@inproceedings{he2019parametric,
  title={Parametric noise injection: Trainable randomness to improve deep neural network robustness against adversarial attack},
  author={He, Zhezhi and Rakin, Adnan Siraj and Fan, Deliang},
  booktitle={Proceedings of the IEEE/CVF Conference on Computer Vision and Pattern Recognition},
  pages={588--597},
  year={2019}
}

@inproceedings{rakin2019bit,
  title={Bit-flip attack: Crushing neural network with progressive bit search},
  author={Rakin, Adnan Siraj and He, Zhezhi and Fan, Deliang},
  booktitle={Proceedings of the IEEE/CVF International Conference on Computer Vision},
  pages={1211--1220},
  year={2019}
}

@inproceedings{criteo-dac,
  title={Criteo Kaggle Display Advertising dataset: https://ailab.criteo.com/ressources}
}

@article{naumov2019deep,
  title={Deep learning recommendation model for personalization and recommendation systems},
  author={Naumov, Maxim and Mudigere, Dheevatsa and Shi, Hao-Jun Michael and Huang, Jianyu and Sundaraman, Narayanan and Park, Jongsoo and Wang, Xiaodong and Gupta, Udit and Wu, Carole-Jean and Azzolini, Alisson G and others},
  journal={arXiv preprint arXiv:1906.00091},
  year={2019}
}

@inproceedings{hochschild2021cores,
  title={Cores that don't count},
  author={Hochschild, Peter H and Turner, Paul and Mogul, Jeffrey C and Govindaraju, Rama and Ranganathan, Parthasarathy and Culler, David E and Vahdat, Amin},
  booktitle={Proceedings of the Workshop on Hot Topics in Operating Systems},
  pages={9--16},
  year={2021}
}

@article{devlin2018bert,
  title={Bert: Pre-training of deep bidirectional transformers for language understanding},
  author={Devlin, Jacob and Chang, Ming-Wei and Lee, Kenton and Toutanova, Kristina},
  journal={arXiv preprint arXiv:1810.04805},
  year={2018}
}
\end{document}